\documentclass{jfm}
\usepackage{graphicx,epstopdf,epsfig,subfig}
\usepackage{amsmath,amssymb,amsfonts,gensymb}
\usepackage{mathrsfs,mathtools,bm,siunitx}
\usepackage{array,booktabs,multicol,multirow}
\usepackage[utf8]{inputenc}
\usepackage[T1]{fontenc}
\usepackage{ulem}
\usepackage{color}
\usepackage{txfonts}
\usepackage{hyperref}

\hypersetup{colorlinks=true,linkcolor=blue,urlcolor=blue,citecolor=blue}
\allowdisplaybreaks[4]

\begin{document}
\shorttitle{Enhancing large-scale motions and turbulent transport in RPPF}
\shortauthor{S. Zhang, Z. Xia, S. Chen}
\title{Enhancing large-scale motions and turbulent transport in rotating plane Poiseuille flow}
\author{Shengqi Zhang\aff{1},
		Zhenhua Xia\aff{2}\corresp{\email{xiazh@zju.edu.cn}} and
		Shiyi Chen\aff{3,1}}
\affiliation{\aff{1} State Key Laboratory for Turbulence and Complex Systems, Peking University, Beijing 100871, China\\
			 \aff{2} Department of Engineering Mechanics, Zhejiang University, Hangzhou 310027, China\\
			 \aff{3} Department of Mechanics and Aerospace Engineering, Southern University of Science and Technology, Shenzhen 518055, China}
\maketitle
\begin{abstract}
Based on the properties of large-scale plume currents, an injection/suction control strategy is introduced to enhance the strength of plume currents and the turbulent transport of passive scalar in rotating plane Poiseuille flow. The control keeps the classical non-penetrative condition on the stable side, and prescribes spanwise varying wall-normal velocity on the unstable side. Direct numerical simulations show that at medium rotation numbers, very weak injection/suction (the root-mean-square velocity on the controlled side is below $1\%$ of the bulk mean velocity) with properly distributed slots could significantly enhance the large-scale plume currents which have the same spatial distribution as the injection slots. In addition, if the distance between plume currents are increased, the turbulent transport efficiency can be promoted. Therefore, taking the strength and transport efficiency into account, the distance between injection slots should be slightly larger than the intrinsic distance between plume currents to achieve maximum enhancement of turbulent transport. For comparison, the control with non-penetrative condition on the unstable side and injection/suction on the stable side is also examined with simulations. It is found that the injection/suction on the stable side has much weaker influence compared with that on the unstable side. This is because that plumes are generated on the unstable side, and that controlling the origin of plumes is more effective.
\end{abstract}
\section{Introduction}\label{sec:intro}
In many geophysical, astrophysical and engineering problems, fluid are bounded by solid walls and affected by system rotation. Such flows show very complex behaviours especially in turbulent state. Rotating plane Poiseuille flow (RPPF) is one of the canonical models of such flows \citep{Johnston1998Effectsof,Jakirlic2002Modelingrotating} for its simplicity, and has been studied intensively for decades with experiments and numerical simulations.

Following the analogy between stratified flows in a gravity field and rotating shear flows \citep{Bradshaw1969Theanalogy}, the flow near the pressure side of RPPF (with higher pressure caused by the Coriolis force corresponding to the mean streamwise velocity) should be destabilized by rotation, and the flow near the suction side (with lower pressure) should be stabilized. Therefore the pressure and suction side are called as unstable and stable side respectively. Turbulent RPPF was first studied by \citet{Johnston1972Effectsof} and it was found that turbulence is enhanced near the unstable side and suppressed near the suction side, which is in accordance with the thermal analogy \citep{Bradshaw1969Theanalogy}. In addition, the local slope of the mean streamwise velocity is found to be twice of the spanwise angular velocity $\Omega_z^*$ approximately, and the Taylor-G\"ortler vortices (roll cells) are also observed. Since heat and mass transfer is important in many industrial apparatus, \citet{Matsubara1996Experimentalstudy} measured the heat and momentum transfer in RPPF, finding that the classical Reynolds analogy is broken by system rotation, and that the heat transfer in RPPF can be almost twice of that in the corresponding non-rotating case. Other experimental studies include but are not limited to \citet{Nakabayashi1996LowReynolds,Nakabayashi2005Turbulencecharacteristics,Maciel200333rdAIAA,Visscher2011Anew}. However, due to the difficulty in the design of experimental apparatus and the global measurement of flow variables, experiments show various limitations in the investigation of RPPF. Instead, direct numerical simulation (DNS) is becoming a prevalent tool for studying RPPF. \citet{Kristoffersen1993Directsimulations} first performed DNS of turbulent RPPF, and observed results consistent with previous experiments. \citet{Nagano2003Directnumerical} simulated scalar transport in RPPF and examined turbulence models. Other DNS studies include but are not limited to \citet{Liu2007Directnumerical,Grundestam2008Directnumerical,Yang2012Channelturbulence,Wallin2013Laminarizationmechanisms,Dai2016Effectsof,Hsieh2016Theminimal,Brethouwer2014Recurrentbursts,Brethouwer2016Linearinstabilities,Brethouwer2017Statisticsand,Brethouwer2018Passivescalar,Brethouwer2019Influenceof,Xia2016Directnumerical,Xia2018Highorder,Zhang2019Atwo}.

In order to improve the performance of industrial apparatus with rotating shear flows, flow control should be considered. \citet{Sumitani1995Directnumerical} studied the influence of homogeneous injection/suction on turbulent friction and heat transport in non-rotating channel flow. They found that injection can increase both the friction coefficient and heat transport, while suction has opposite effects. Also in channel flow, \citet{Choi1994Activeturbulence} introduced the active control which applies the injection/suction with the opposite wall-normal velocity of that measured at a certain distance from the wall, and observed a reduction of skin friction up to $25\%$ in DNS. \citet{Iuso2002Wallturbulence} controlled streamwise vortices in turbulent channel flow with vortex generator jets, and achieved the reduction of local and global skin friction up to $30\%$ and $15\%$ respectively. \citet{Bakhuis2020Controllingsecondary} studied the influence of spanwise-varying wall roughness on large-scale structures and total drag of Taylor-Couette flow (TCF), finding that proper distribution of roughness can rearrange large-scale vortices. \citet{Zhang2020Controllingflow,Zhang2021Stabilizingdestabilizing} introduced side wall temperature control to Rayleigh-B\'enard convection (RBC) to fix the separation points of plumes on side walls, which can enhance the large-scale circulation and heat transport. There are rudimentary studies on flow control for RPPF. \citet{Younis2012Predictionof} used Reynolds-averaged equations to estimate the effect of homogeneous injection/suction on RPPF. However, the results are not verified by experiments or DNS, and homogeneous injection/suction does not seem realizable in industry. \citet{Wu2019Turbulencestatistics} studied the influence of homogeneous wall roughness on RPPF, finding that roughness on the stable side can destabilize the flow and counteracts the stabilizing effect of rotation. Nevertheless, the studies on flow control for RPPF is not enough. As shown in our previous work, RPPF has plume-like structures (called as plumes) which form large-scale plume currents. However, different from classical TCF which has plumes emitting from both inner and outer walls, RPPF only has the unstable side as the source of plumes. This certainly means that the control on the unstable side and the stable side should have very different effects. Therefore, it is necessary to examine the performance of flow control on each side of RPPF separately.

In this paper, we will introduce an injection/suction control strategy for turbulent RPPF and examine the performance through DNS. The numerical set-up of RPPF under injection/suction control will be introduced in \S\ref{sec:setup}. The simulation results of large-scale plume currents and turbulent scalar transport are demonstrated and discussed in \S\ref{sec:results}. Finally, the present work will be summarized in \S\ref{sec:conclusion}.
\section{Governing equations and numerical description}\label{sec:setup}
\subsection{Governing equations and boundary conditions}\label{subsec:setup_equation}

\begin{figure}
	\centering
	\includegraphics[width=0.6\textwidth]{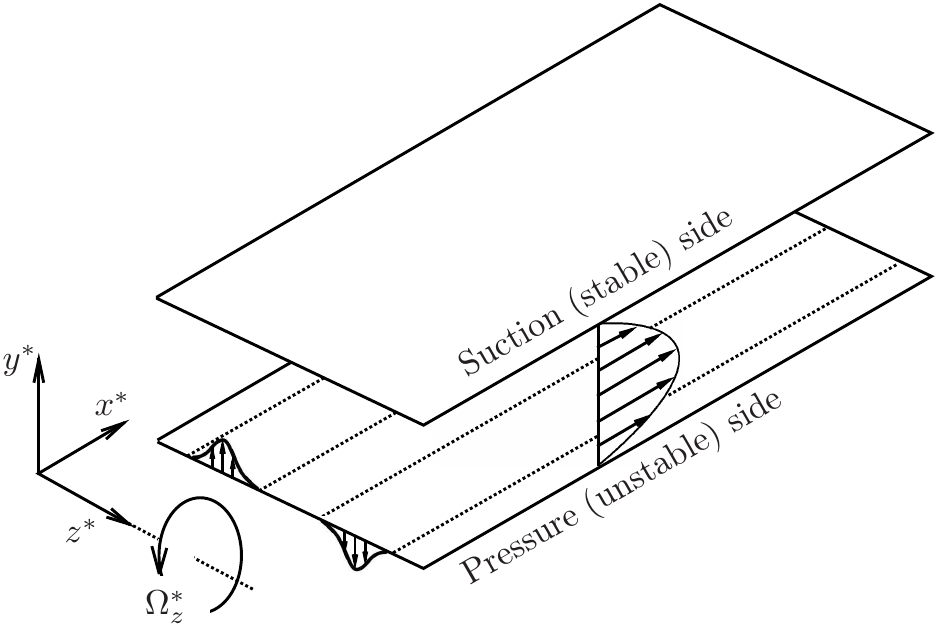}
	\caption{Sketch of the three-dimensional spanwise rotating plane Poiseuille flow.}
	\label{fig:sketch_3D_psfrag}
\end{figure}

As sketched in figure~\ref{fig:sketch_3D_psfrag}, the incompressible fluid is bounded by two infinite parallel plates rotating along the spanwise ($z$) direction with angular velocity $\Omega_z^*$. The reference frame also rotates following the plates, so that the vertical coordinates ($y$) of the plates could be fixed at $y^*=\pm h^*$ with $h^*$ being the channel half width. The fluid is driven by a constant body force $dP^*/dx^*$ in the streamwise ($x$) direction. By convention, the lower/upper wall is called as the pressure (unstable)/suction (stable) side when $\Omega_z^*>0$. A passive scalar $\phi^*$ has constant values $\phi_B^*$ and $\phi_T^*$ at the lower and upper walls respectively. The constant reference density is $\rho^*$; the kinematic viscosity is $\nu^*$; the diffusivity of $\phi^*$ is $\kappa^*$. After choosing reference length scale $h^*$, reference passive scalar scale $\phi_B^*-\phi_T^*$ and reference velocity scale $u_\tau^*=\sqrt{-dP^*/dx^*(h^*/\rho^*)}$, the governing equations and boundary conditions can be written in a non-dimensionalized form:
\begin{equation}\begin{split}
	\nabla\cdot\bm{u}&=0,\\
	\frac{\partial\bm{u}}{\partial t}+\bm{u}\cdot\nabla\bm{u}
		=-\nabla p+&Re_\tau^{-1}\nabla^2\bm{u}-Ro_\tau\bm{e}_z\times\bm{u},\\
	\frac{\partial\phi}{\partial t}+\bm{u}\cdot\nabla\phi=&Pr^{-1}Re_\tau^{-1}\nabla^2\phi,\\
	y=-1:\ \phi=1,\ \bm{u}=-V_B&\bm{e}_y\sum\limits_{k=-\infty}^{\infty}(-1)^k\eta\left(\frac{z-kD_J}{d_J}\right),\\
	y=+1:\ \phi=0,\ \bm{u}=+V_T&\bm{e}_y\sum\limits_{k=-\infty}^{\infty}(-1)^k\eta\left(\frac{z-kD_J}{d_J}\right),
\end{split}\label{eqn:3D}\end{equation}
with global friction Reynolds number $Re_\tau=u_\tau^*h^*/\nu^*$, global friction rotation number $Ro_\tau=2\Omega_z^*h^*/u_\tau^*$, and Prandtl number $Pr=\nu^*/\kappa^*$. In the velocity boundary condition, $V_B$ and $V_T$ are maximum wall-normal velocity at the lower and upper walls respectively, $D_J$ is the spanwise distance between the centrelines of two neighbouring slots, $d_J$ is the slot width, and
\begin{equation}
	\eta(\xi)=\left\{\begin{split}
	\cos^2&(\pi\xi),\ &|\xi|\leq\frac{1}{2}\\
	&0,\ &|\xi|>\frac{1}{2}
	\end{split}\right.
\end{equation}
is the basic distribution function of wall-normal velocity of each slot.
\subsection{Numerical set-up}\label{subsec:setup_numerical}
\begin{figure}
	\centering
	\includegraphics[width=0.5\textwidth]{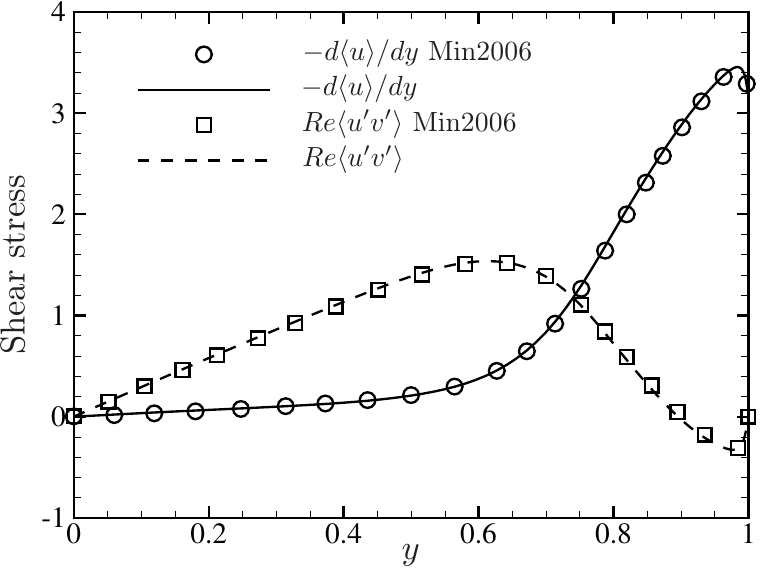}
	\caption{Viscous and Reynolds shear stresses of the present validation case and those in \citet{Min2006Sustainedsub}.}
	\label{fig:Jet_AFiD_Valid_psfrag}
\end{figure}

The system is simulated with a second-order central-difference code AFiD \citep{VanDerPoel2015Apencil} with modifications. Flow variables are defined in a rectangular domain periodic in horizontal directions, and discretized on a staggered grid. The pressure Poisson equation is solved using discrete Fourier transform in horizontal directions and a tridiagonal solver. The explicit second-order Adams-Bashforth scheme is chosen for time marching. Since AFiD is designed for simulating Rayleigh-B\'enard convection, the $\phi$ equation could be solved well. A validation case without rotation is simulated with parameters and boundary conditions:
\begin{equation}\begin{split}
	Re_b=1333.3,\ v(x,\pm1,&z)=\pm a\cos(\alpha(x-ct))\\
	a=0.15U_b,\ \alpha=&0.5,\ c=-3U_b
\end{split}\label{eqn:Validation}\end{equation}
where $U_b$ being the bulk mean velocity and $Re_b=Re_\tau U_b$. Figure~\ref{fig:Jet_AFiD_Valid_psfrag} shows the viscous and Reynolds shear stresses computed using the present code, and they match well with the results in \citet{Min2006Sustainedsub}.

In all following simulations, the spanwise period $L_x=4\pi$ and streamwise period $L_z=2\pi$. Other basic parameters are shown in table~\ref{tab:AllPara}. $N_J$ is the number of injection slots in a spanwise period and $D_J=L_z/N_J$. $V_B$ and $V_T$ are chosen so that the root-mean-square (r.m.s.) of $v$ at the injection/suction wall is $0.1$ ($0.1u_\tau$). For cases with $Ro_\tau>0$, injection/suction are applied on either side of the walls, while for $Ro_\tau=0$ the control is only applied on the lower wall due to symmetry.

\begin{table}\begin{center}
		\begin{tabular}{m{1.4cm}<{\centering}		m{0.8cm}<{\centering}		m{0.8cm}<{\centering}		m{0.8cm}<{\centering}		m{0.8cm}<{\centering}		m{0.8cm}<{\centering}		m{0.8cm}<{\centering}		m{0.8cm}<{\centering}		m{0.8cm}<{\centering}		m{1.4cm}<{\centering}		m{0.8cm}<{\centering}	}
			\toprule																					
			Case	&	$Re_\tau$	&	$Ro_\tau$	&	$Pr$	&	$V_B$	&	$V_T$	&	$N_J$	&	$d_J$	&	$\Delta_x^+$	&	$\Delta_y^+$	&	$\Delta_z^+$	\\
			\midrule																					
			R0J0	&	$180$	&	$0$	&	$1$	&	$0$	&	$0$	&	-	&	-	&	$8.84$	&	$[0.52,2.46]$	&	$4.42$	\\
			R0J2B	&	$180$	&	$0$	&	$1$	&	$0.327$	&	$0$	&	$2$	&	$\pi/8$	&	$8.84$	&	$[0.52,2.46]$	&	$4.42$	\\
			R0J3B	&	$180$	&	$0$	&	$1$	&	$0.267$	&	$0$	&	$3$	&	$\pi/8$	&	$8.84$	&	$[0.52,2.46]$	&	$4.42$	\\
			R0J4B	&	$180$	&	$0$	&	$1$	&	$0.231$	&	$0$	&	$4$	&	$\pi/8$	&	$8.84$	&	$[0.52,2.46]$	&	$4.42$	\\ \specialrule{0.1pt}{0pt}{1.5pt}
			R5J0	&	$180$	&	$5$	&	$1$	&	$0$	&	$0$	&	-	&	-	&	$8.84$	&	$[0.52,2.46]$	&	$4.42$	\\
			R5J2B	&	$180$	&	$5$	&	$1$	&	$0.327$	&	$0$	&	$2$	&	$\pi/8$	&	$8.84$	&	$[0.52,2.46]$	&	$4.42$	\\
			R5J2T	&	$180$	&	$5$	&	$1$	&	$0$	&	$0.327$	&	$2$	&	$\pi/8$	&	$8.84$	&	$[0.52,2.46]$	&	$4.42$	\\
			R5J3B	&	$180$	&	$5$	&	$1$	&	$0.267$	&	$0$	&	$3$	&	$\pi/8$	&	$8.84$	&	$[0.52,2.46]$	&	$4.42$	\\
			R5J3T	&	$180$	&	$5$	&	$1$	&	$0$	&	$0.267$	&	$3$	&	$\pi/8$	&	$8.84$	&	$[0.52,2.46]$	&	$4.42$	\\
			R5J4B	&	$180$	&	$5$	&	$1$	&	$0.231$	&	$0$	&	$4$	&	$\pi/8$	&	$8.84$	&	$[0.52,2.46]$	&	$4.42$	\\
			R5J4T	&	$180$	&	$5$	&	$1$	&	$0$	&	$0.231$	&	$4$	&	$\pi/8$	&	$8.84$	&	$[0.52,2.46]$	&	$4.42$	\\ \specialrule{0.1pt}{0pt}{1.5pt}
			R10J0	&	$180$	&	$10$	&	$1$	&	$0$	&	$0$	&	-	&	-	&	$8.84$	&	$[0.52,2.46]$	&	$4.42$	\\
			R10J2B	&	$180$	&	$10$	&	$1$	&	$0.327$	&	$0$	&	$2$	&	$\pi/8$	&	$8.84$	&	$[0.52,2.46]$	&	$4.42$	\\
			R10J2T	&	$180$	&	$10$	&	$1$	&	$0$	&	$0.327$	&	$2$	&	$\pi/8$	&	$8.84$	&	$[0.52,2.46]$	&	$4.42$	\\
			R10J3B	&	$180$	&	$10$	&	$1$	&	$0.267$	&	$0$	&	$3$	&	$\pi/8$	&	$8.84$	&	$[0.52,2.46]$	&	$4.42$	\\
			R10J3T	&	$180$	&	$10$	&	$1$	&	$0$	&	$0.267$	&	$3$	&	$\pi/8$	&	$8.84$	&	$[0.52,2.46]$	&	$4.42$	\\
			R10J4B	&	$180$	&	$10$	&	$1$	&	$0.231$	&	$0$	&	$4$	&	$\pi/8$	&	$8.84$	&	$[0.52,2.46]$	&	$4.42$	\\
			R10J4T	&	$180$	&	$10$	&	$1$	&	$0$	&	$0.231$	&	$4$	&	$\pi/8$	&	$8.84$	&	$[0.52,2.46]$	&	$4.42$	\\ \specialrule{0.1pt}{0pt}{1.5pt}
			R30J0	&	$180$	&	$30$	&	$1$	&	$0$	&	$0$	&	-	&	-	&	$8.84$	&	$[0.52,2.46]$	&	$4.42$	\\
			R30J2B	&	$180$	&	$30$	&	$1$	&	$0.327$	&	$0$	&	$2$	&	$\pi/8$	&	$8.84$	&	$[0.52,2.46]$	&	$4.42$	\\
			R30J2T	&	$180$	&	$30$	&	$1$	&	$0$	&	$0.327$	&	$2$	&	$\pi/8$	&	$8.84$	&	$[0.52,2.46]$	&	$4.42$	\\
			R30J3B	&	$180$	&	$30$	&	$1$	&	$0.267$	&	$0$	&	$3$	&	$\pi/8$	&	$8.84$	&	$[0.52,2.46]$	&	$4.42$	\\
			R30J3T	&	$180$	&	$30$	&	$1$	&	$0$	&	$0.267$	&	$3$	&	$\pi/8$	&	$8.84$	&	$[0.52,2.46]$	&	$4.42$	\\
			R30J4B	&	$180$	&	$30$	&	$1$	&	$0.231$	&	$0$	&	$4$	&	$\pi/8$	&	$8.84$	&	$[0.52,2.46]$	&	$4.42$	\\
			R30J4T	&	$180$	&	$30$	&	$1$	&	$0$	&	$0.231$	&	$4$	&	$\pi/8$	&	$8.84$	&	$[0.52,2.46]$	&	$4.42$	\\
			\bottomrule																					
		\end{tabular}
\end{center}
\caption{Basic physical and computational parameters. The wall units are defined using $u_\tau^*$.}
\label{tab:AllPara}\end{table}
\section{Results}\label{sec:results}
\subsection{Large-scale motions}\label{subsec:results_motions}
\begin{figure}
	\centering
	\includegraphics[width=1\textwidth]{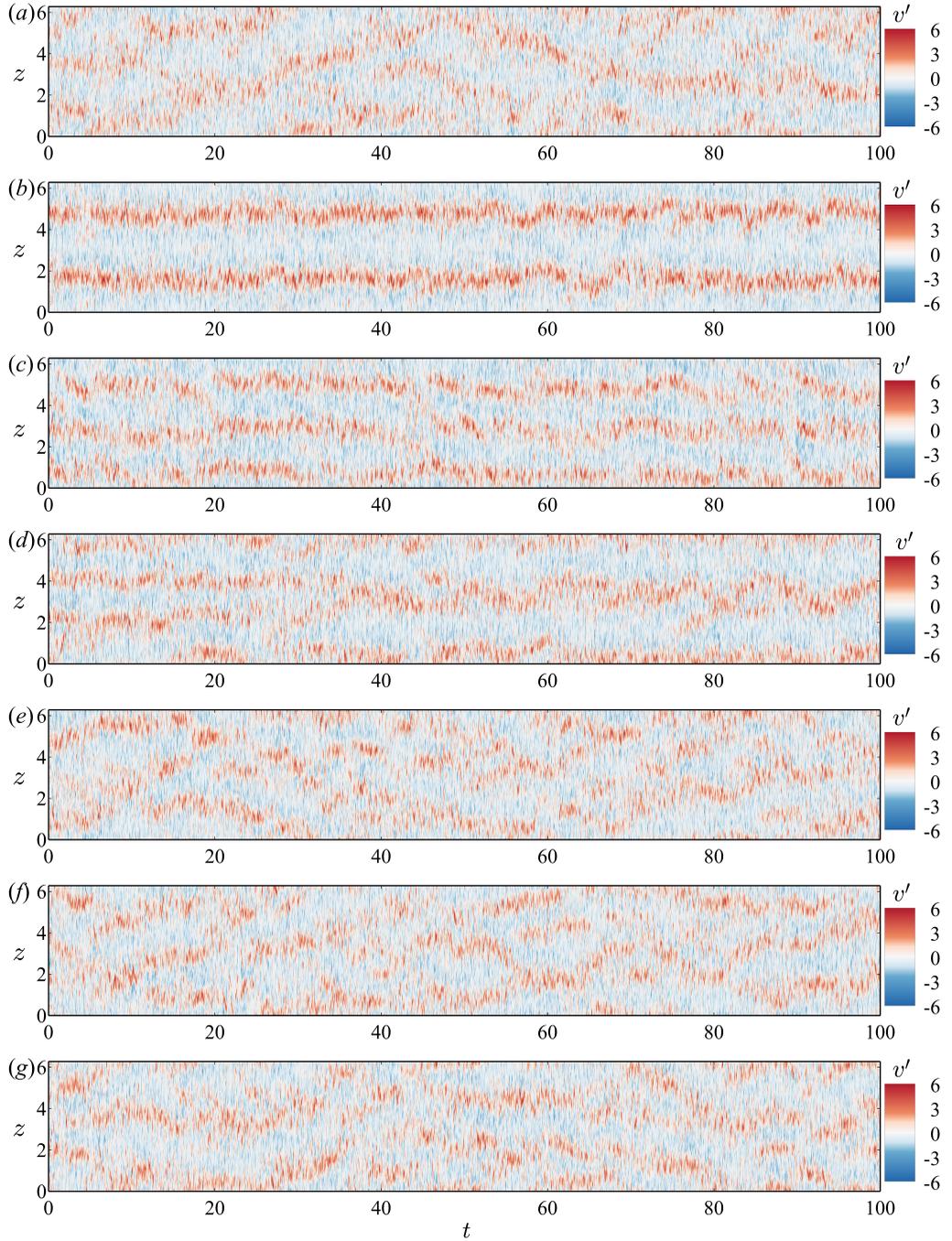}
	\caption{Contours of $v'(0,0,z,t)$ in the $Ro_\tau=5$ cases. The origin of $t$ is chosen within a statistically steady state. (a) R5J0; (b) R5J2B; (c) R5J3B; (d) R5J4B; (e) R5J2T; (f) R5J3T; (g) R5J4T.}
	\label{fig:V_y0_R5Jet}
\end{figure}

\begin{figure}
	\centering
	\includegraphics[width=1\textwidth]{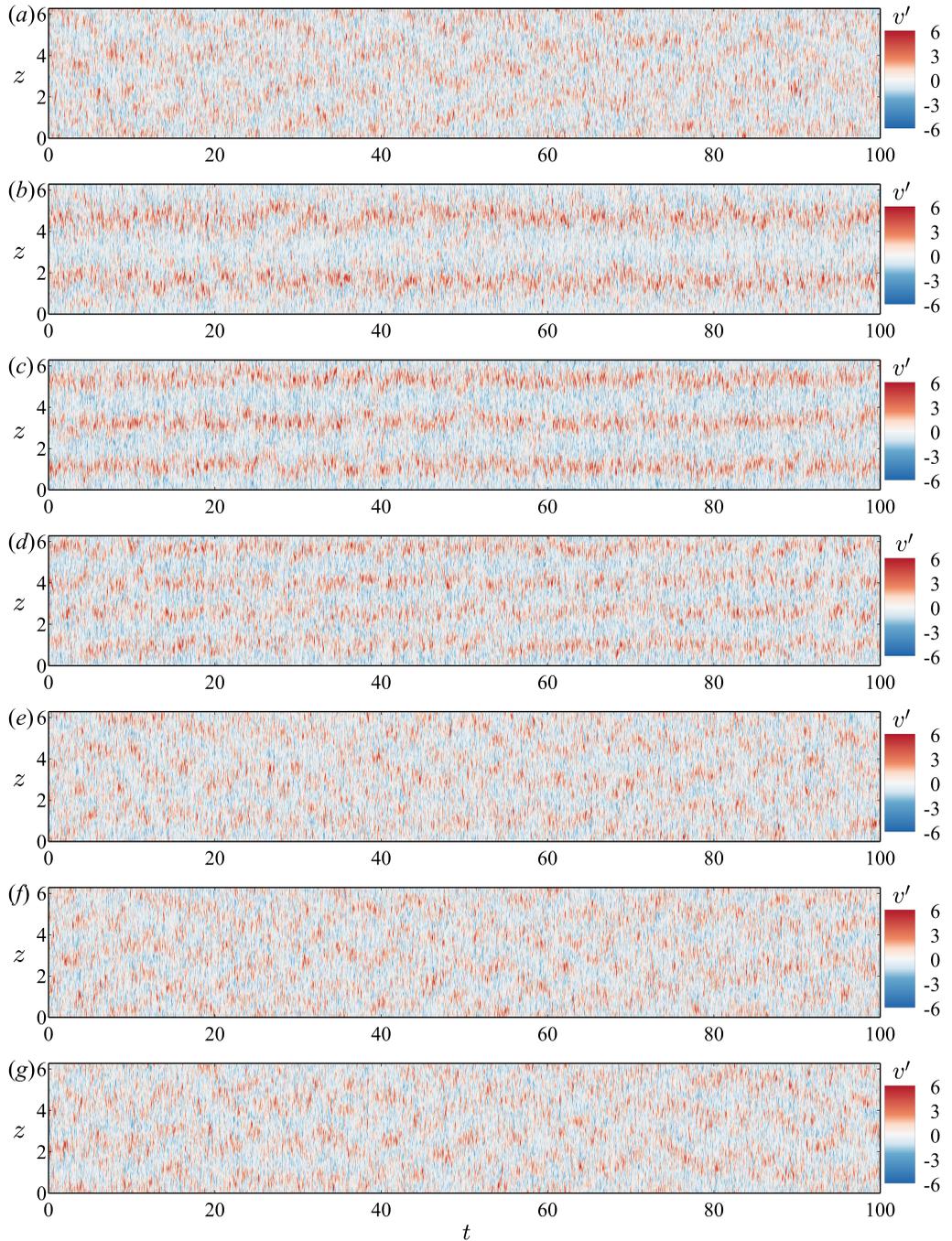}
	\caption{Contours of $v'(0,0,z,t)$ in the $Ro_\tau=10$ cases. The origin of $t$ is chosen within a statistically steady state. (a) R10J0; (b) R10J2B; (c) R10J3B; (d) R10J4B; (e) R10J2T; (f) R10J3T; (g) R10J4T.}
	\label{fig:V_y0_R10Jet}
\end{figure}

As shown in our previous work, plume currents formed by rising plumes generated from the unstable side are the most important large-scale structures in RPPF. For a certain small or medium $Ro_\tau$, the number and pattern of plume currents are constantly changing with $t$. Figure~\ref{fig:V_y0_R5Jet} shows the contours of $v'$ in the $t-z$ plane with $x=y=0$ at $Ro_\tau=5$, which intuitively demonstrates the patterns and dynamic behaviours of plume currents. It can be clearly seen from figure~\ref{fig:V_y0_R5Jet}(a) that there can be two or more plume currents in the uncontrolled case, and they are constantly splitting, merging and oscillating. With the control of two (R5J2B) and three (R5J3B) injection slots (in a spanwise period) on the unstable side, the number of plume currents (in a spanwise period) is fixed at two and three respectively, as shown in figure~\ref{fig:V_y0_R5Jet}(b,c). Furthermore, the plume currents are oscillating near the injection slots. It is worth mentioning that the injection/suction velocity is much smaller than the wall-normal velocity in plume currents (the root-mean-square wall-normal velocity on the unstable side is about $0.6\%$ of the bulk mean velocity). The main reason for the strong influence of the relatively weak control is that plumes separate from the unstable side and the injection induces the plumes to separate near the injection slots, which is similar as the control strategy introduced in \citet{Zhang2020Controllingflow,Zhang2021Stabilizingdestabilizing}. However, with four injection slots (R5J4B), the number of plume currents cannot be fixed and can be two, three and four. What is different from the uncontrolled case is that the splitting and merging of plume currents are relatively slower, and the spanwise locations of plume currents are often close to injection slots or in the middle of two injection slows. A plume current in the middle of two injection slots are likely to be the result of two neighbouring plume currents which have merged. The relatively smaller influence of four injection slots is probably due to the fact that the intrinsic spanwise distance between plume currents is around $\pi$ and $2\pi/3$ in RPPF at $Re_\tau=180$ and $Ro_\tau=5$. Therefore four plume currents (with average distance $\pi/2$) induced by four injection slots tend to merge. Figure~\ref{fig:V_y0_R10Jet} shows the contours of $v(0,0,z,t)'$ at $Ro_\tau=10$. It can be indicated from figure~\ref{fig:V_y0_R5Jet}(a) and figure~\ref{fig:V_y0_R10Jet}(a) that the distances between plume currents are generally smaller at $Ro_\tau=10$, which is explained in our previous work. As a result, four injection slots applied to the unstable side will induce four relatively stable plume currents in the $Ro_\tau=10$ case, since the control pattern is close to the intrinsic pattern of plume currents.

For comparison, the behaviours of plume currents under the injection/suction control on the stable side are shown in figure~\ref{fig:V_y0_R5Jet}(e-g) and figure~\ref{fig:V_y0_R10Jet}(e-g). Apparently the large-scale motions include splitting, merging and spanwise oscillation, and their activeness seem to be the same as those in the uncontrolled case. This indicates that such control have no significant influence on large-scale motions. The main reason is that the separation of plumes on the unstable side do not strongly depend on what is happening on the stable side, which should be the common property in both RPPF and penetrative convection.


\begin{figure}
	\centering
	\includegraphics[width=1\textwidth]{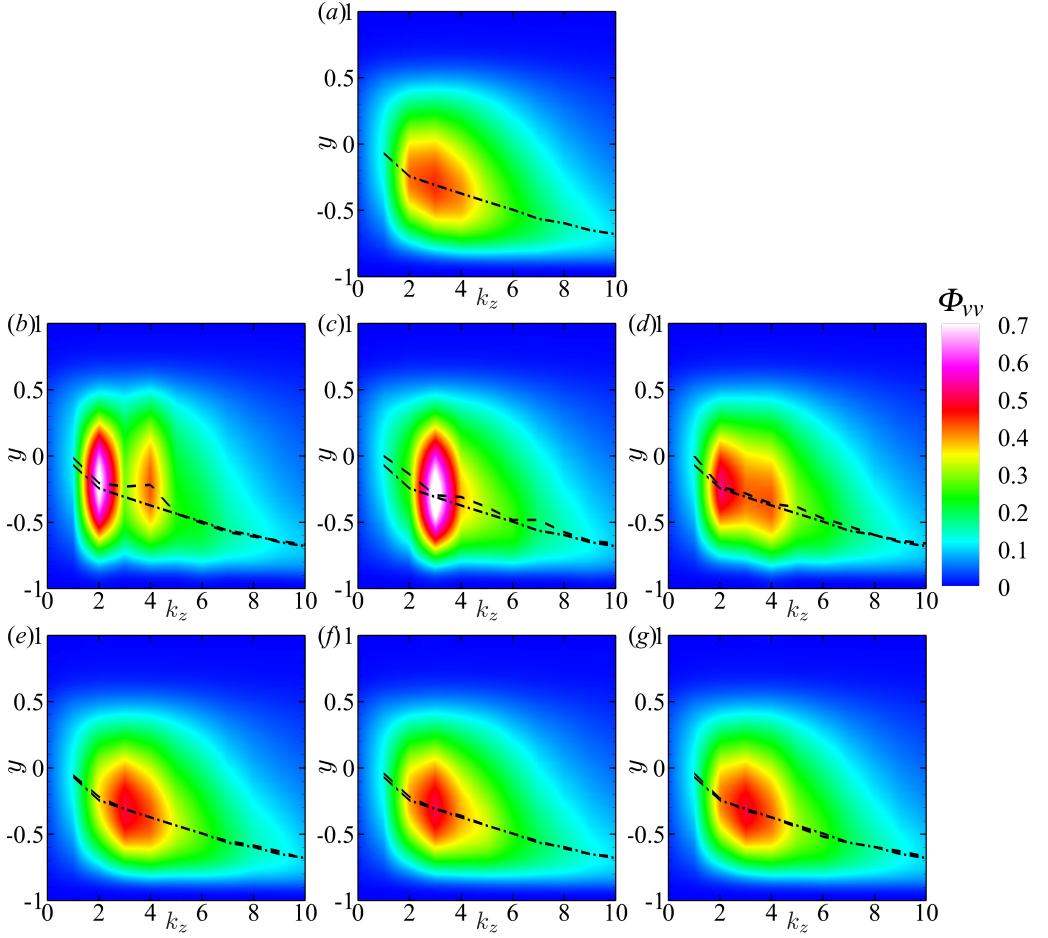}
	\caption{Contour of spanwise energy spectra of wall-normal velocity $\varPhi_{vv}(k_z;y)$ in the $Ro_\tau=5$ cases. (a) R5J0; (b) R5J2B; (c) R5J3B; (d) R5J4B; (e) R5J2T; (f) R5J3T; (g) R5J4T.  Dashed lines denote the vertical locations of largest $\varPhi_{vv}(k_z;y)$ at each $k_z$ of the corresponding cases, and dash-dotted lines denote those of the uncontrolled case R5J0.}
	\label{fig:Spec_ykz_Jet_R5}
\end{figure}

Figure~\ref{fig:Spec_ykz_Jet_R5} shows the spanwise energy spectra $\varPhi_{vv}(k_z;y)$ of $v'$ (horizontal Fourier energy spectra summed in $k_x$ direction). Apparently the injection/suction is so weak that it hardly has any direct influence on $\varPhi_{vv}$ near the walls, again demonstrating the sensitivity of large-scale motions to the control configuration. In the uncontrolled case R5J0, the height of the maximum point of a spanwise mode decreases with $k_z$ when $k_z\leq10$. This is because plume currents with height much larger than spanwise distance will be unstable and tend to merge with neighbouring ones (also explained in our previous work). However, figure~\ref{fig:Spec_ykz_Jet_R5}(b) shows that with two injection slots on the unstable side, the maximum point of $k_z=4$ mode is close to that of $k_z=2$ mode. The main reason is that the state with two stable plume currents is not sinusoidal in the $z$ direction, and will induce a new $k_z=4$ mode different from that corresponding to four plume currents (in R5J0). Figure~\ref{fig:Spec_ykz_Jet_R5}(c) indicates that the three plume currents induced by three injection slots on the unstable side is closer to sinusoidal and mainly contains $k_z=3$ mode. Figure~\ref{fig:Spec_ykz_Jet_R5}(d) shows that four injection slots on the unstable side only enhances the $k_z=2$ and $k_z=4$ mode, without changing their maximum points. This is because the $k_z=2$ mode is mainly caused by the merging of plume currents above injection slots, and should have similar behaviour as the original $k_z=2$ mode in the uncontrolled case. Figure~\ref{fig:Spec_ykz_Jet_R5}(e-g) again illustrates that the weak injection/suction on the stable side does not strongly change the behaviour of large-scale motions.

\begin{figure}
	\centering
	\includegraphics[width=0.9\textwidth]{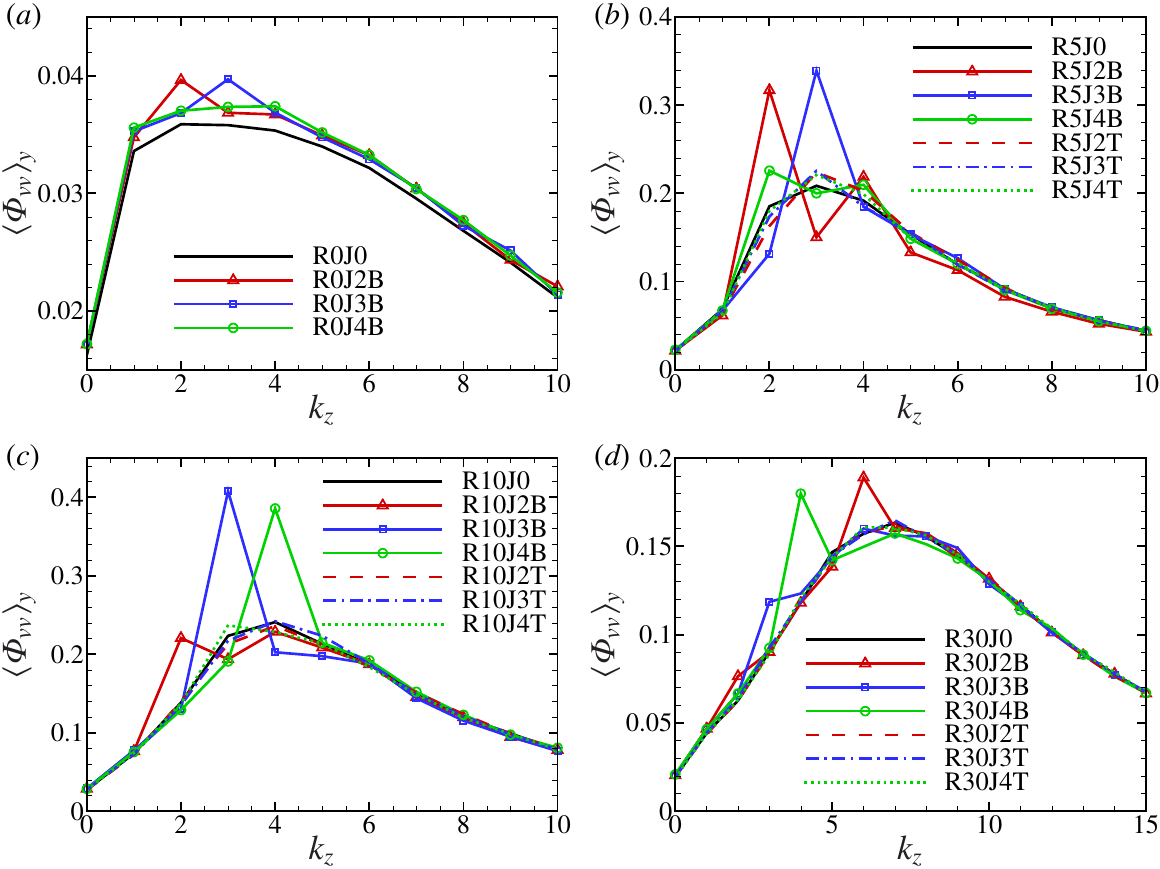}
	\caption{$\varPhi_{vv}(k_z;y)$ averaged over $y$. (a) $Ro_\tau=0$; (b) $Ro_\tau=5$; (c) $Ro_\tau=10$; (d) $Ro_\tau=30$.}
	\label{fig:Spec_kz_Jet}
\end{figure}

Figure~\ref{fig:Spec_kz_Jet} shows $\langle\varPhi_{vv}\rangle_y$, which is the spanwise $v'$ energy spectra $\varPhi_{vv}(k_z;y)$ averaged along the $y$ direction. Here $\langle\cdot\rangle$ with subscript $x$, $y$, $z$ or $t$ represents the average in the corresponding direction. It can be seen that the large-scale modes of the uncontrolled $Ro_\tau=0$ case is much weaker than those of the $Ro_\tau=5,10,30$ cases. In addition, the plane Poiseuille flow without rotation does not have the similar plume dynamics like RPPF. Therefore, the maximum enhancement of the $k_z\in[1,8]$ modes in $Ro_\tau=0$ cases with injection/suction control is small compared with RPPF cases with injection/suction control on the unstable side. Figure~\ref{fig:Spec_kz_Jet}(b,c) indicates a more detailed explanation for the discrepancy between the control configuration and the plume current distribution in R5J4B. It is likely that the dominating mode of the uncontrolled case represent the distribution of plume currents that is most efficient in collecting plumes (totally). In other words, the distance between neighbouring plume currents with the dominating pattern tends to be as small as possible (close to the limit of stability), so that each plume is close enough to a plume current and get pulled towards it. Therefore, a pattern of plume currents (four plume currents induced by four injection slots at $Ro_\tau=5$) that is more compact than the dominating pattern in the uncontrolled case ($k_z=3$ mode at $Ro_\tau=5$) is likely to exceed the critical height-over-distance ratio of stability and tends to merge. Another interesting phenomenon can be found in figure~\ref{fig:Spec_kz_Jet}(c) corresponding to $Ro_\tau=10$, which shows that the control with three and four injection slots on the unstable side can induce large increase in the energy of $k_z=3$ and $k_z=4$ modes respectively, while the increase of the $k_z=2$ mode caused by two injection slots is relatively small. This indicates that, a control pattern which is too sparse compared with the dominating large-scale pattern in the uncontrolled case, would be less capable in enhancing large-scale motions. The underlying mechanism should be that a sparse pattern of plume currents may fail to collect some near-wall plumes. Those plumes will be dissipated by viscosity rapidly or form new plume currents. Such mechanism is further supported by figure~\ref{fig:Spec_kz_Jet}(d), corresponding to $Ro_\tau=30$ at which the $k_z=7$ mode is dominating in the uncontrolled case. It can be seen that three injection slots on the unstable side is less effective than four injection slots. In addition, two injection slots on the unstable side mainly promotes the $k_z=6$ mode, indicating that there are more plume currents between injection slots.

The discussion above can be summarized into simple rules of the present injection/suction control strategy to enhance large-scale motions in RPPF. At a fixed $Re_\tau$ and $Ro_\tau$, $n$ injection slots on the unstable side (in $z\in[0,2\pi]$) with $n$ not larger than and not much smaller than the $k_z$ of the dominating large-scale pattern of the uncontrolled case can significantly enhance the plume currents with the corresponding pattern. Too compact injection slots will make plume currents to merge, and too sparse injection slots will waste the energy of some plumes or induce new plume currents. Therefore too compact or too dense control pattern with $n$ injection slots on the unstable side in $z\in[0,2\pi]$ will not significantly enhance the $k_z=n$ mode. The injection/suction control on the stable side can barely enhance large-scale motions because plumes do not originate or separate on the stable side.

\subsection{Turbulent transport}\label{subsec:results_transport}
Following the definition in \citet{Brethouwer2018Passivescalar} and equation~\eqref{eqn:3D}, the Nusselt number of $\phi$ has equivalent definitions (assuming that the averaging time is infinite)
\begin{equation}
	Nu=\left.-2\frac{\partial\langle\phi\rangle}{\partial y}\right|_{y=-1}
	=\left.-2\frac{\partial\langle\phi\rangle}{\partial y}\right|_{y=1}
	=1+PrRe_\tau\int_{-1}^1{\langle v'\phi'\rangle dy}.
\end{equation}
Here, $\langle\cdot\rangle$ represents the average in $x$, $z$ and $t$ directions. Apparently, increasing the turbulent transport $\langle v'\phi'\rangle$ is the only way to increase the scalar transport since the contribution of molecular diffusion is fixed.

\begin{figure}
	\centering
	\includegraphics[width=0.9\textwidth]{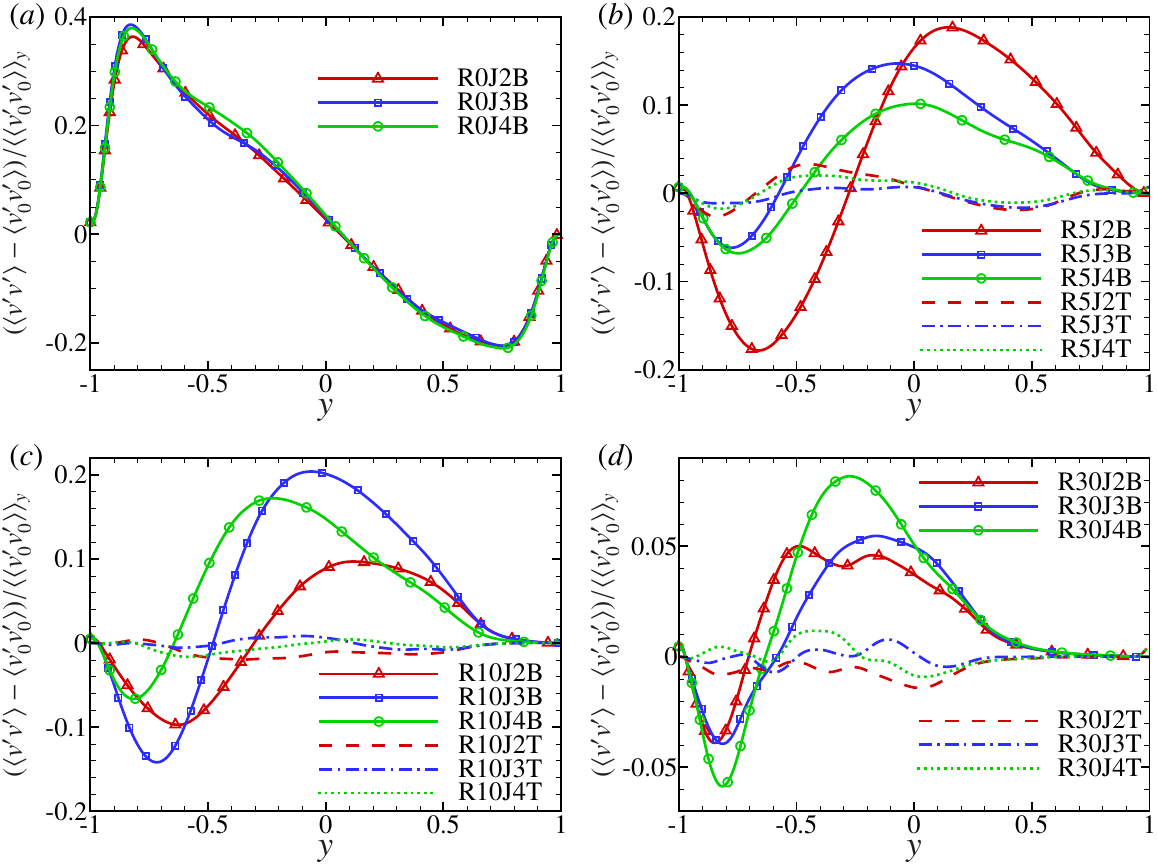}
	\caption{Relative deviation of $\langle v'v'\rangle$ from that of the corresponding uncontrolled cases. (a) $Ro_\tau=0$; (b) $Ro_\tau=5$; (c) $Ro_\tau=10$; (d) $Ro_\tau=30$.}
	\label{fig:Stat_dvv_Jet}
\end{figure}

\begin{figure}
	\centering
	\includegraphics[width=0.9\textwidth]{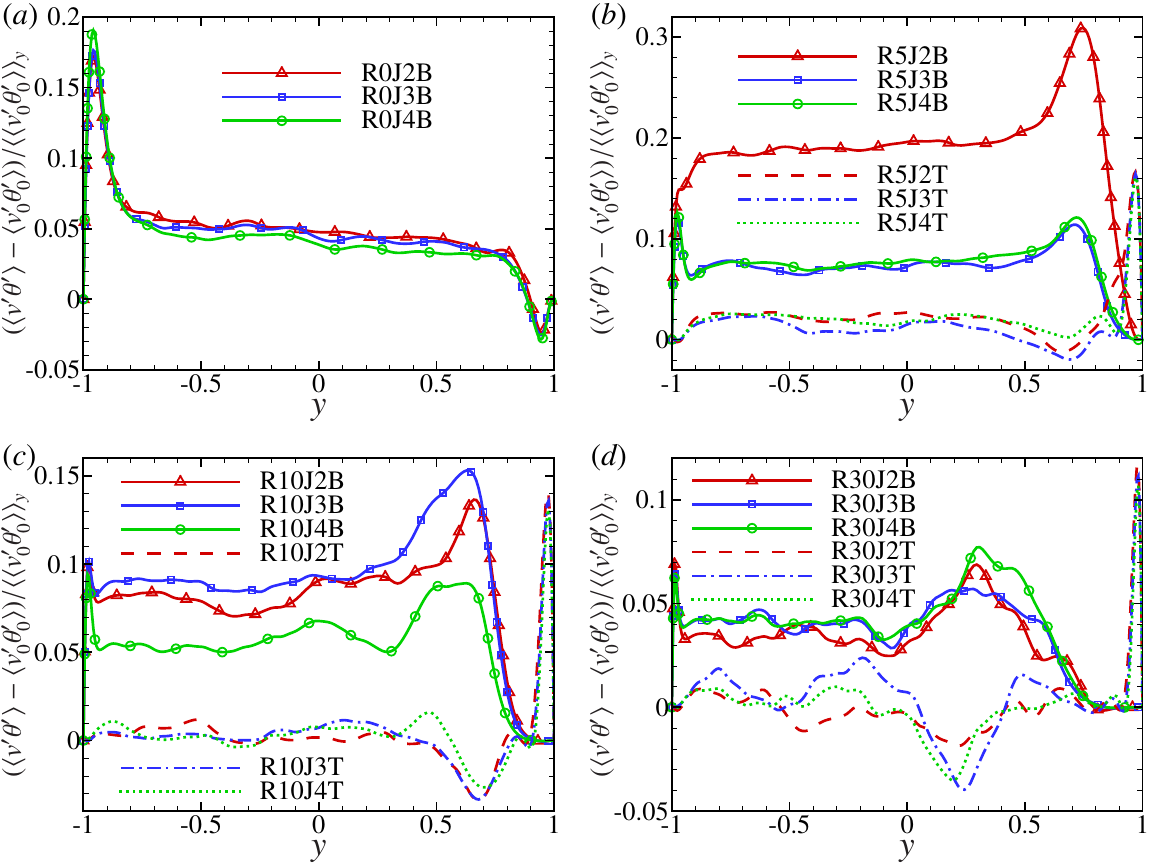}
	\caption{Relative deviation of $\langle v'\phi'\rangle$ from that of the corresponding uncontrolled cases. (a) $Ro_\tau=0$; (b) $Ro_\tau=5$; (c) $Ro_\tau=10$; (d) $Ro_\tau=30$.}
	\label{fig:Stat_dvq_Jet}
\end{figure}

Figure~\ref{fig:Stat_dvv_Jet} and figure~\ref{fig:Stat_dvq_Jet} show respectively the relative enhancement of $\langle v'v'\rangle$ and $\langle v'\theta'\rangle$ achieved by the present control. It should be mentioned that $\langle v'v'\rangle$ is fixed at $0.01$ on the injection/suction wall. Therefore, figure~\ref{fig:Stat_dvv_Jet}(a) and~\ref{fig:Stat_dvq_Jet}(a) indicate that the increase of $\langle v'v'\rangle$ and $\langle v'\theta'\rangle$ does not strongly depend on the distribution of injection/suction slots, and should be mainly determined by the $\langle v'v'\rangle$ on the controlled wall. It is probably because that the turbulent motions that contribute to $\langle v'v'\rangle$ and scalar transport in non-rotating Poiseuille flow are dominated by small-scale structures, and not sensitive to the large-scale distribution of injection/suction slots. When $Ro_\tau$ is non-zero and not large, the distribution of injection/suction slots on the unstable side becomes very important to the kinetic energy of wall-normal velocity and turbulent scalar transport. As shown in figure~\ref{fig:Stat_dvq_Jet}(b), two injection slots on the unstable side induces the largest enhancement of $\langle v'\theta'\rangle$. There should be two main reasons. The first is that case R5J2B has a large increase of large-scale kinetic energy close to that of case R5J3B, as shown in figure~\ref{fig:Spec_kz_Jet}(b) and figure~\ref{fig:Stat_dvv_Jet}(b). The second is that two plume currents have larger spanwise distance than three plume currents, and larger distance between plume currents may lead to higher efficiency of scalar transport as mentioned in our previous work. However, when $Ro_\tau=10$, three injection slots on the unstable side has a better performance than two injection slots as shown in figure~\ref{fig:Stat_dvq_Jet}(c). This is because that $\langle v'v'\rangle$ in case R10J3B is generally larger than that in case R10J2B as shown in figure~\ref{fig:Stat_dvv_Jet}(c) and the higher efficiency of two plume currents in scalar transport cannot compensate their lower strength. When $Ro_\tau=30$, the increase of $\langle v'v'\rangle$ and $\langle v'\theta'\rangle$ induced by injection/suction on the unstable side is relatively small compared with the corresponding cases at $Ro_\tau=5$ or $Ro_\tau=10$. This suggests that turbulent kinetic energy and scalar transport is sensitive to the control only when the scale of control pattern and the intrinsic scale of large-scale motions are close. The injection/suction on the stable side scarcely affects $\langle v'v'\rangle$ and $\langle v'\phi'\rangle$ as shown in figure~\ref{fig:Stat_dvv_Jet}(b,c,d), except near the stable side. A reason in addition to its small influence on plume currents is that, inside the region with $d\langle u\rangle/dy-Ro_\tau\ll0$ all vertical motions with frequency $0$ are damped by rotation, as shown in our previous work.

Therefore, a promising control strategy to promote turbulent scalar transport in RPPF should be applying injection/suction on the unstable wall. The distance between injection slots should be slightly smaller than the dominating spanwise wavelength $\lambda_z=2\pi/k_z$ in the uncontrolled case in order to guarantee both the strength of plume currents and the efficiency of scalar transport.

\section{Conclusion}\label{sec:conclusion}
In this paper, a control configuration for enhancing large-scale motions and turbulent scalar transport is introduced. If the distance between injection slots is not smaller and not much larger than the dominating spanwise wavelength of plume currents in the uncontrolled case, the control on the unstable side can adjust the spanwise distance of plume currents and greatly increase their strength. The underlying mechanism is that injection can force plumes on the unstable side to separate near the injection slots and avoid the splitting or merging of plume currents. When the spanwise distance between plume currents in the controlled case is slightly larger than that of the uncontrolled case, turbulent transport of passive scalar can be significantly enhanced due to the enhancement of both strength and transport efficiency of plume currents. The same injection/suction control applied to the stable side has very weak influence on plume currents and scalar transport.
\section*{Acknowledgements}
This work was supported by the National Science Foundation of China (NSFC grant nos. 11822208, 11988102, 11772297, 91852205) and Shenzhen Science and Technology Program (Grant No. KQTD20180411143441009). The numerical simulations were finished at National Supercomputer Center in Guangzhou (Tianhe-2A), China.

\section*{Declaration of interests}
The authors report no conflict of interest.
\bibliographystyle{jfm}
\bibliography{Reference}

\end{document}